\documentclass[prd,twocolumn]{revtex4}
\usepackage{amsmath}
  \newif\ifpdf \ifx\pdfoutput\undefined \pdffalse \else \pdftrue \fi 
\ifpdf \usepackage[pdftex]{graphics} \else \usepackage{graphics} \fi 
\usepackage{graphicx}
\usepackage{epsfig}
\usepackage{color}

\newcommand {\abs}[1]{\left| #1 \right|}

\begin{document}

\def \Vista {{\sc Vista}}
\def \Sleuth {{\sc Sleuth}}
\def \Quaero {{\sc Quaero}}
\def \met {{\,/\!\!\!\!E_{T}}}
\def \scriptP {{\ensuremath{\cal P}}}
\def \scriptR {{\ensuremath{\cal R}}}
\def \twiddleScriptP {{\ensuremath{\tilde{\cal P}}}}
\def \xvec {{\ensuremath{\vec{x}}}}
\def \figuresize {5.5in}
\newcounter{mylistc}
\newenvironment{mylist}
        {\setcounter{mylistc}{0}
         \begin{list}{$\bullet$}
        {\usecounter{mylistc}
         \setlength{\parsep}{0pt}
         \setlength{\itemsep}{0pt}}}{\end{list}}

\title{Statistical Challenges with Massive Data Sets in Particle Physics}
\author{Bruce Knuteson}
\homepage{http://hep.uchicago.edu/~knuteson/}
\email{knuteson@fnal.gov}
\affiliation{Enrico Fermi Institute, University of Chicago}
\author{Paul Padley}
\homepage{http://www-d0.fnal.gov/~padley/}
\email{padley@rice.edu}
\affiliation{Rice University}

\date{\today}

\begin{abstract}

The massive data sets from today's particle physics experiments present a variety of challenges amenable to the tools developed by the statistics community.  From the real-time decision of what subset of data to record on permanent storage, to the reduction of millions of channels of electronics to a few dozen high-level variables of primary interest, to the interpretation of these high-level observables in the context of an underlying physical theory, there are many problems that could benefit from a higher-bandwidth exchange of ideas between our fields.  Examples of interesting problems from various stages in the collection and interpretation of particle physics data are provided in an attempt to whet the appetite of future collaborators with knowledge of potentially helpful techniques, and to encourage fruitful discussion between the particle physics and statistics communities.

\end{abstract}

\maketitle

\tableofcontents

\section{Introduction\label{sec:Introduction}}

\begin{figure}[htb]
\includegraphics[angle=90,width=3.5in]{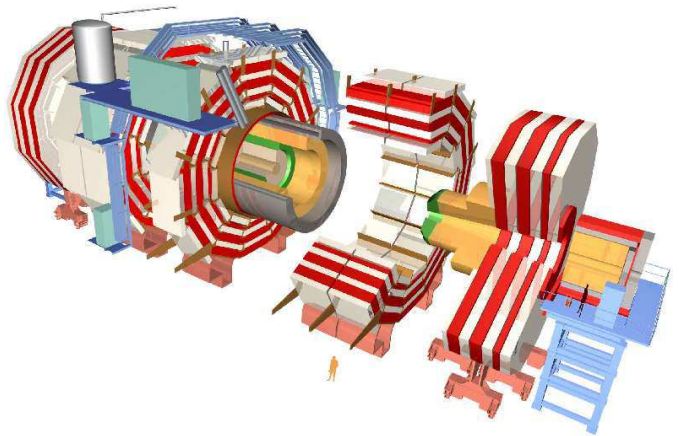} 
\caption{A cut away view of the CMS detector, currently under construction.  Note the person included for scale.  {\tiny{(Credit: the CMS Collaboration)}}} 
\label{fig:CMS}
\end{figure}

Particle physics is the science focused on the understanding of the most elementary constituents of matter and the forces governing their interactions.  In stark contrast to the minute size of the objects under investigation, enormous experiments are required to further this understanding.  A typical next generation experiment --- the Compact Muon Solenoid (CMS)~\cite{cmsTDR} at the Large Hadron Collider (LHC), located at the European Organization for Nuclear Research (CERN)~\cite{CERN} 
near Geneva, Switzerland --- will weigh 12.5 thousand tonnes and have millions of channels of electronics, producing nearly 40 terabytes of data per second.  A cartoon of the CMS detector is provided in Fig.~\ref{fig:CMS}.  These data will be analyzed in real time and reduced to approximately ten terabytes per day that will be stored for later analysis.  Of the one billion particle collisions occurring each second at the LHC, only a few are of interest.  Finding these interesting --- but possibly unanticipated --- collisions in such a massive data stream represents a challenging test of forefront technology and computational algorithms.

Particle physics data come in discrete packets referred to as {\em events}.  In the case of a proton collider experiment, an event is a head-on collision of two protons, each moving at nearly the speed of light.  The energy of the two colliding protons can allow the creation of new matter, by Einstein's $E = m c^2$, which is spewed forth as debris from the collision.  Particle physics detectors are designed to record and identify this debris.

\begin{figure}[htb]
\includegraphics[width=3.5in]{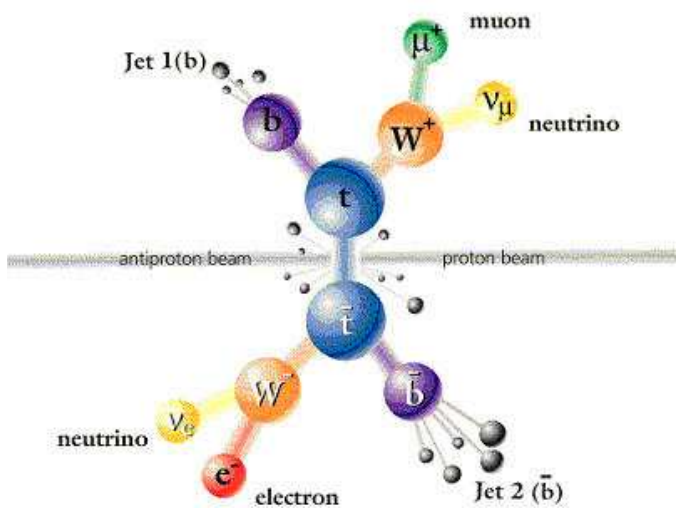}
\caption{A cartoon of an interesting proton anti-proton collision.  In this event two top quarks ($t$ and $\bar{t}$) are produced, which decay in roughly $10^{-24}$ seconds to six particles, here labeled $b$, $\mu^+$, $\nu_\mu$, $\bar{b}$, $e^-$, and $\nu_e$.  The energies and directions of these final state particles are (imperfectly) measured or inferred by the surrounding detector.  {\tiny{(Credit: Fermilab)}}} \label{fig:CartoonFinalState} \end{figure}

Sections~\ref{sec:Triggering} and~\ref{sec:Reconstruction} describe the real-time reduction of the massive data stream emerging from particle physics detectors to a handful of the most important components of those debris, each carrying a direction and energy.  The total number of observables in the final state after this massive reconstruction is roughly two dozen.  A cartoon characterization of these observables is presented in Fig.~\ref{fig:CartoonFinalState}. 

\begin{figure}[htb]
\includegraphics[width=3.5in]{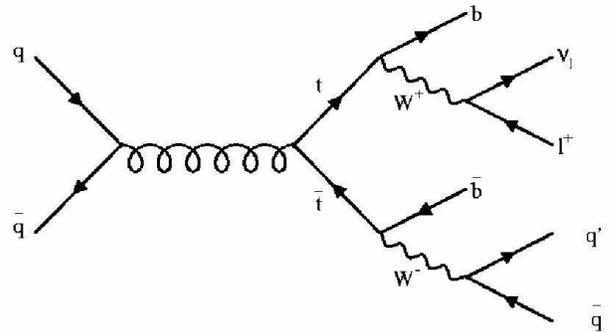}
\caption{A Feynman diagram.  Time flows to the right.  This diagram depicts a quark ($q$) and an anti-quark ($\bar{q}$) colliding, producing two top quarks ($t$ and $\bar{t}$) that decay through $W$ bosons ($W^+$ and $W^-$) into six particles ($b$, $\nu_l$, $l^+$, $\bar{b}$, $q'$, and $\bar{q}$).}
\label{fig:ttbarFeynmanGraph}
\end{figure}

Sections~\ref{sec:KernelDensityEstimation} and~\ref{sec:Sleuth} describe two methods that use the resulting two dozen high-level observables to connect to the underlying physical theory.  Our present understanding of that underlying model of particle physics is encapsulated in the quantum mechanical ``Standard Model,'' which predicts (probabilistically) the debris that will emerge.  In practice the calculations are typically done using {\em Feynman diagrams}, named after the late Caltech professor, which serve as both a convenient calculational tool and an intuitive graphical aid.  An example of such a diagram is shown in Fig.~\ref{fig:ttbarFeynmanGraph}.  A set of {\em Feynman rules} associates each piece of the diagram --- each vertex and each line --- with an algebraic expression, which after some manipulation allows a prediction to be made for the fraction of collisions that will produce certain types of debris.  

The statistical analysis of these data occurs at many levels and in various forms.  This paper sketches a few of the challenges currently faced in particle physics for which statistical methods are being employed, emphasizing examples from pattern recognition and signal identification.  The ultimate challenge is to cull interesting and unanticipated events from billions of collisions in a data stream flowing at terabytes per second. 

\section{Triggering\label{sec:Triggering}}

If the detector components that record the debris of each collision are thought of as the sense organs of the detector, the {\em trigger} should be thought of as a hardwired reflex.  Many animals are programmed to respond to quick motion in their peripheral vision; analogously, today's collider detectors are designed to respond to events with certain characteristics.  While motion in an animal's periphery may take hundreds of milliseconds to trigger the appropriate muscular response for identifying or fleeing from an apparent threat, the experiments at the LHC must decide 40 million times per second whether the collisions that have occurred are sufficiently interesting to be worth recording permanently.  

The large volume of data produced in today's particle physics detectors precludes a full analysis of all events; it is not even physically possible to record so much data.  A fast analysis of the data is therefore performed in real time at various levels of detail, and data deemed sufficiently interesting are saved for a more thorough subsequent analysis.

\subsection{Trigger levels}

Many of today's particle physics experiments employ a trigger system with three levels:
\begin{itemize}
\item{{\bf Level 1}. Custom electronics located physically on the detector perform a fast analysis within the detector components themselves, reducing the data that is subsequently passed through communication networks downstream.}
\item{{\bf Level 2}. Detector components are read out into computers that perform a regional analysis of the data.}
\item{{\bf Level 3}. Data from different regions of the detector are brought together in a single CPU to allow a final global decision.}
\end{itemize}

An animal's response to a perceived threat is determined by a neural system that can be placed somewhere between ``hardware'' (built-in, unmodifiable features that come with the biological package) and ``software'' (modifiable logic learned as the organism develops).  Level 1 of many particle physics triggers make heavy use of ``firmware,'' often in the form of field programmable gate arrays (FPGAs).  These devices combine the hardware advantage of high operational speeds with the flexibility and reprogrammability of software, allowing physicists to change the algorithms used as experience is gained.

The Level 2 trigger uses the longer decision time available to it to combine information from several sub-detectors, and to perform a more sophisticated selection of the data than is possible at Level 1.   In cases for which two bytes of information are available from a particular detector element, the Level 1 trigger may use only the coarse information available in the first byte; at Level 2 the full information may be used.  Calibration corrections that are not available in Level 1 can be applied to the data at Level 2.  It is also possible at this stage to consider multivariate analysis methods.

At Level 3, the entire event is available for consideration.  At this stage a fast version of the final analysis can be performed, and time-intensive algorithms can be brought to bear.  Level 3 typically is allocated $\approx 1$~second of decision time per event.
 
\subsection{Example}

\begin{figure}
\resizebox{3.5in}{!}{\hskip -1.2cm \includegraphics{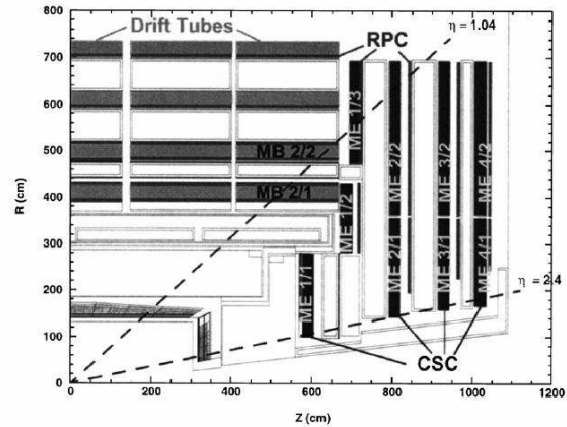} }
\caption{A cross sectional view of a quarter of the CMS detector.  The colliding beams run along the axis defined by $R=0$, with collisions occurring at $(z,R)=(0,0)$.  The end cap muon chambers are shown, separated by slabs of steel over a meter thick.  Muons are unique in that they can penetrate this steel.  {\tiny{(Credit: the CMS Collaboration)}}}
\label{fig:Endcap}
\end{figure}

The end cap muon trigger for the CMS experiment~\cite{Acosta:2003fv} is an example of a practical trigger architecture.  This trigger identifies muon particles traveling through the end caps of the detector~\cite{cmsMuonTDR}, the signature for which is a charged particle traveling through the entire apparatus.  In the end cap region there are chambers, labeled CSC's in Fig.~\ref{fig:Endcap}, which can capture the ionization trail of the particle.

\begin{figure}
\resizebox{2.5in}{!}{\hskip -1.0cm \includegraphics{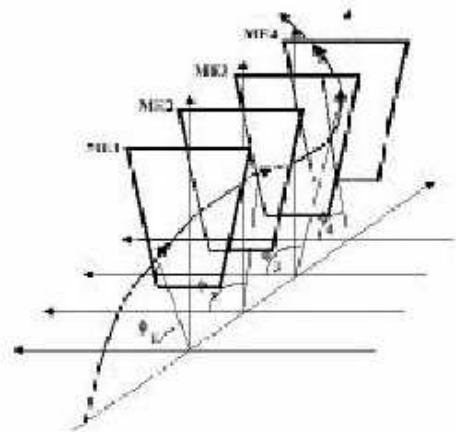} } 
\caption{A sketch of the curved path a muon could take through the end cap muon system.}
\label{fig:muonTrack}
\end{figure}

As can be seen in Fig.~\ref{fig:muonTrack}, a muon takes a complicated path through the chambers, due to the varying magnetic field in the steel structure in which the chambers are embedded.   The Level 1 trigger's goal is to select muons that follow paths through the detector that correspond to high momentum.  Low momentum muons, which exhibit more bending as they go through the chambers, are of lesser interest.  There is a probability distribution of possible bend angles and paths through the detector for a muon of given momentum.  The trigger uses these distributions, generated by simulation, to select events that are likely to contain high momentum muons.  The two key enabling technologies for this effort have general applicability to the real time analysis of massive streams of data: sorting in field programmable gate arrays, and the use of memory lookups.

\begin{figure}
\resizebox{3.5in}{!}{\hskip -0.5cm \includegraphics[angle=0]{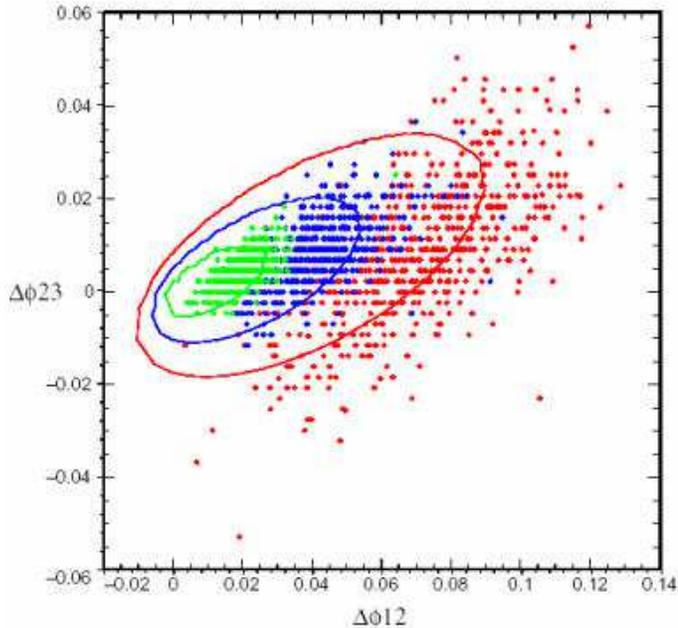} } 
\caption{A scatter plot of the $\Delta\phi_{12}$ and $\Delta\phi_{23}$ variables for 
simulated muons of $p_T=$ 3 (red), 5 (blue), and 10 (green) GeV.  Overlaid are contours of the function $p_T(\Delta\phi_{12},\Delta\phi_{23})$, obtained by maximizing the likelihood function in Eq.~\ref{eqn:endcapMuonLikelihood}.}
\label{fig:DeltaPhi}
\end{figure}

Field programmable gate arrays are large scale chips, comprising many gates and memory cells, that can be configured to carry out custom algorithms.  Gate arrays with a million logic units are not uncommon these days, and a custom CPU can be built by appropriately configuring the gates.  In designing high-speed analyses of massive data streams, algorithms involving loops are to be avoided at all costs.  Using FPGAs implementing massively parallel algorithms, a printed circuit board has been built that can sort in real time the nearly 30 gigabytes of data per second incident to it as part of the CMS end cap muon trigger~\cite{Matveev:2001mk}.   Such techniques have become commonplace in experimental particle physics, and have been used in a variety of triggers.

A second key enabling technology involves memory lookup devices.  In the CMS end cap muon system, the muon's momentum between two chambers can be found as a function of the difference in bend angle in those chambers,
$$p_T = \frac{A_{ij}}{\Delta \phi_{ij}},$$
where $A_{ij}$ is determined by fitting the results of a simulation, and allowed to vary with position in the detector.  Maximization of the likelihood 
\begin{widetext} 
\begin{equation}
\label{eqn:endcapMuonLikelihood}
{\cal L }= \frac{1}{2\pi\sigma_{12}\sigma_{23}\sqrt{1-\rho^2}}
\exp\left(\frac{-1}{2(1-\rho^2)}\left[\frac{(\Delta\phi_{12}-\mu_{12})^2}{\sigma^2_{12}}
- \frac{2\rho(\Delta\phi_{12}-\mu_{12})(\Delta\phi_{23}-\mu_{23})}{\sigma_{12}\sigma_{23}}
+ \frac{(\Delta\phi_{23}-\mu_{23})^2}{\sigma^2_{23}}\right]\right)
\end{equation} 
\end{widetext}
as a function of $p_T$ is desired, where $\mu_{ij}=A_{ij}/p_T$, $\sigma_{ij}$ is the error in $\Delta\phi_{ij}$, and it is assumed that the ratios $\sigma_{ij}/\mu_{ij}$ are independent of $p_T$.  The result is shown in Fig.~\ref{fig:DeltaPhi}.  The mapping between measured angle differences $\Delta\phi_{12}$ and $\Delta\phi_{23}$ and the most likely momentum $p_T$ is then loaded into a memory chip.  This enables the calculation of a fairly complex function in the time it takes to access the memory, in this case about 12.5 nanoseconds~\cite{Wang:2001gk}.  The computational and graphical statistics community could make welcome contributions to the determination of suitable functions for memory lookups in similar problems.  

\section{Reconstruction\label{sec:Reconstruction}}

For each one-in-a-million collision selected by the trigger, the primary components of the debris spewed forth in the collision are reconstructed.  This process is analogous to the investigation of a head-on automobile collision, noting the resultant trajectories of broken doors, shattered glass, and strewn metal.  The two examples considered here are track finding and jet clustering.

\subsection{Track Finding}

\begin{figure}
\resizebox{3.5in}{!}{\hskip -0.5cm \includegraphics{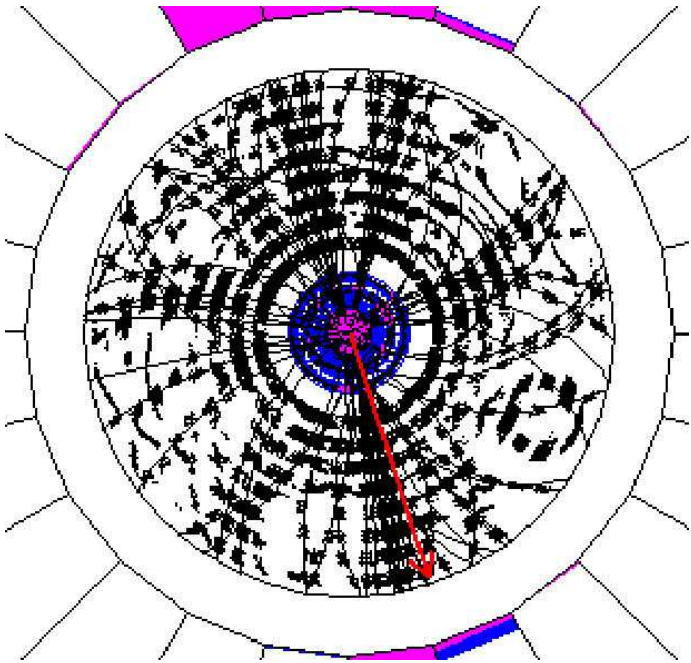} } 
\caption{End view of a collision in the roughly cylindrical CDF detector.  Small black dots indicate points through which a charged particle has passed, identified in the detector.  These ``hits'' are sewn together to reconstruct probable particle trajectories, shown by curves emanating from the collision at the center of the figure.  The magnetic field lines flow perpendicular to the page, causing the tracks to curve by an amount proportional to the strength of the magnetic field and inversely proportional to the momentum of the particle.  {\tiny{(Credit: the CDF Collaboration)}}}
\label{fig:Tracking}
\end{figure}

Many of the particles produced in an event are electrically charged.  These can be detected by any of several technologies, including the trail of ionization left in a volume of gas, the electron-hole pairs liberated in a semiconductor, and the scintillation light emitted upon excitation of certain complex organic molecules.  In all cases the detector is trying to measure in three dimensions the passage of charged particles, usually by combining three two-dimensional views at three different angles through the detector.  Tracking systems are typically embedded in a magnetic field, which bends the track by an amount inversely proportional to the momentum of the particle perpendicular to the field, allowing a measurement of this momentum.

In Fig.~\ref{fig:Tracking}, the small dots represent the measured points of passage of charged particles.  The superimposed lines are fits to the tracks.  There are two stages to making these fits: pattern recognition, in which points that represent a common track are linked together; and fitting, in which track parameters, such as the momentum of the particle and its point of closest approach to the collision point, are extracted.  The process is complicated by the presence of spurious noise hits, measurement error, and possible change of the particle's direction through interaction with material.  The determination of charged particle trajectories can be the most time consuming part of the reconstruction of an event; new, more efficient methods of track finding and fitting are continuously being sought.

A typical approach to track fitting involves finding a set of hit points that represent a possible track, forming a $\chi^2$ quantifying how well the measured points fit particular track parameters, and then minimizing this $\chi^2$ with respect to those parameters.  Correlations among measurements, introduced for example by the interaction of the particle with the material it is traversing, complicates and slows the inversion of the covariance matrix required for this minimization.  The Kalman filter~\cite{KalmanFilter} is commonly used in practice.

\subsection{Jet Clustering}

\begin{figure}[ht]
\resizebox{3.5in}{!}{\hskip -0.5cm \includegraphics[angle=0]{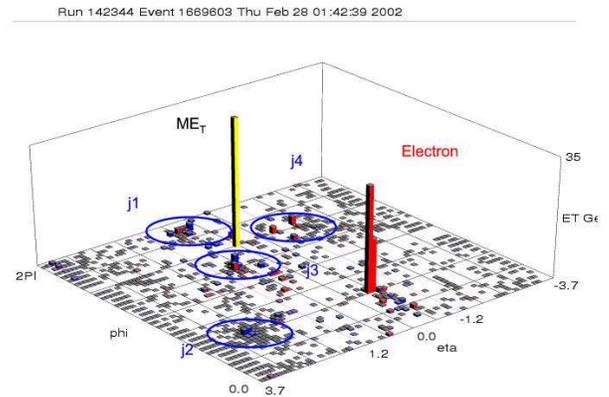} } 
\caption{An example of jet clustering.  A rolled out view of the detector is shown.  The bin heights correspond to energy deposited in cells of the calorimeter.  Clusters of energy in the detector are associated with jets, shown by the circles enclosing them.  {\tiny{(Credit: the D\O\ Collaboration)}}} 
\label{fig:jetClustering}
\end{figure}

A second common reconstruction problem in particle physics is jet clustering in calorimeters, detectors that measure the energy of particles entering them.  An energetic, strongly-interacting particle entering a calorimeter causes a chain reaction of nuclear breakup and particle production, resulting in a shower of particles passing through the detector.  Figure~\ref{fig:jetClustering} shows a calorimeter unfolded out into a two-dimensional array; the height of each cell is proportional to the amount of energy deposited in it in this particular event.  Quarks produced in the original collision fragment into many particles, which appear in the detector as a collimated ``jet'' of energy flow.  The pattern recognition problem involves identifying the clumps of energy in the calorimeter that constitute a jet, allowing an estimate of the energy and direction of the parent quark.  Less-than-perfect theoretical understanding of the fragmentation process complicates this already messy problem.

There are other sources of jets that one would like to identify.  For example, $\tau$ leptons can decay into a tightly collimated jet of energy in the calorimeter.  While the statistical distribution of energy from a $\tau$ decay is different on average from that obtained from the fragmentation of quarks, there is a significant overlap in the distributions.  Optimal use of measured information to best separate jets from quarks and jets from $\tau$ leptons is an outstanding problem in proton collider experiments.
 
\section{Kernel Density Estimation\label{sec:KernelDensityEstimation}}

Section~\ref{sec:Triggering} discussed how potentially interesting events are triggered on in real time; Sec.~\ref{sec:Reconstruction} sketched how the millions of channels of detector information characterizing each event are reduced to roughly two dozen numbers describing the trajectories of the primary components of the debris from each collision.  This section and the next summarize two methods designed to facilitate the interpretation of these events in the context of an underlying physical theory.

The problem of interpretation can often be formulated as a classification or discrimination problem, with the goal being to distinguish between two different hypotheses.  Our understanding of the underlying physics is such that one of the two hypotheses is invariably the Standard Model, and the second is often an extension of the Standard Model.  This leads naturally into thinking of the problem as searching for a ``signal'' over a ``background.''  Here and below the signal will be denoted by $S$, and the background will be denoted by $B$.  The two hypotheses being compared are then schematically $A=S+B$ versus $B$.  Particle physicists often perform this comparison in a rather simplistic way, carving out one contiguous region in a low-dimensional variable space, counting the number of events seen in the data within that region, and comparing to the number of events predicted from the two hypotheses $A$ and $B$ under consideration.  

The modeling of the response of our detector to this debris is done through a rather detailed Monte Carlo simulation of each detector subcomponent.  In the simulation the debris is tracked through detector components, causing signals on individual channels in the same way that real debris produces such signals in our physical detectors.  The simulated events are then processed using the same algorithm as was used on the data.  The simulation of these events in the detector often requires substantial time, on the order of several seconds per event on a modern (at the time of this writing) 2 GHz CPU.  This limits the number of events available to model the predictions from different hypotheses of the underlying physics.  In nearly all cases, time and other practical constraints limit the number of simulated events with which we can populate the space of two dozen observables to on the order of one million.

\subsection{The problem}

The issue described in this section could be avoided if we had an analytic form for the theoretical prediction in our space of observables.  The Monte Carlo nature of our simulation prevents this; some prescription for smoothing out the density of points obtained from this Monte Carlo approach is therefore required.  Even one million events do not adequately populate a ten-dimensional space, so physicists are forced to consider a subset of relevant variables for testing hypotheses of interest.  The nature of each hypothesis to be tested typically immediately suggests a variable space ${\cal V}$ of low dimensionality that is apt to be most useful for distinguishing the hypothesis from the Standard Model, which has been the default hypothesis in the field for the past two decades.  The dimensionality of ${\cal V}$ is typically between one and four, each dimension representing a simple algebraic combination of the original two dozen observables.  In practice, the choice of which variables to use to test a given hypothesis is often a matter of much consternation and discussion.

\subsection{Kernel solution}

There are a number of ways to determine the region in which to do this counting.  One such procedure, still used to a surprising extent, is to eyeball the predicted distributions for the two hypotheses and choose some rectangular box within the variable space by considering one-dimensional projections.  A more sophisticated option, and one enjoying increasing popularity among particle physicists, is to use a neural network trained to distinguish between the two different hypotheses.  

A third option, and the one we advocate, is to use kernels to estimate probability densities for the two hypotheses within the variable space ${\cal V}$.  The simplest estimate gives the probability densities $p(\vec{x}|S)$ and $p(\vec{x}|B)$ at a point $\vec{x}$ within ${\cal V}$ in terms of the Monte Carlo points ${\vec x}_{Si}$ and ${\vec x}_{Bj}$ in ${\cal V}$, each with weight $w_{Si}$ and $w_{Bj}$, drawn from the probabilistic predictions of $S$ and $B$, as \begin{widetext}
\begin{equation}
p(\vec{x}|S) = \frac{1}{\sum_i{w_{Si}}}{\sum_i{\frac{w_{Si}}{\sqrt{(2\pi)^d\abs{\Sigma_S}} {h_S}^d}e^{\left(-(\xvec-\xvec_{Si})^T \Sigma_S^{-1} (\xvec-\xvec_{Si})/2 {h_S}^{2d}\right)}}}
\end{equation}
and
\begin{equation}
p(\vec{x}|B) = \frac{1}{\sum_j{w_{Bj}}}{\sum_j{\frac{w_{Bj}}{\sqrt{(2\pi)^d\abs{\Sigma_B}} {h_B}^d}e^{\left(-(\xvec-\xvec_{Bj})^T \Sigma_B^{-1} (\xvec-\xvec_{Bj})/2 {h_B}^{2d}\right)}}}.
\end{equation} \end{widetext}
Here $\vec{x}_{Si}$ denotes the position within the $d$-dimensional variable space ${\cal V}$ of the $i^{\text{th}}$ Monte Carlo event used to model hypothesis $S$, and $\vec{x}_{Bj}$ denotes the position of the $j^{\text{th}}$ Monte Carlo event used to model hypothesis $B$.  The width of the Gaussian kernels are set by the covariance matrices $\Sigma_S$ and $\Sigma_B$ of the points $\{\xvec_{Si}\}$ and $\{\xvec_{Bj}\}$, and by smoothing parameters $h_S$ and $h_B$ that depend upon the number of Monte Carlo events available and the dimensionality of ${\cal V}$. 

The two densities $p(\vec{x}|S)$ and $p(\vec{x}|B)$ are then used to define a discriminant
\begin{equation}
\label{eqn:D}
D(\vec{x}) = \frac{p(x|S)}{p(x|S) + p(x|B)}.
\end{equation}
The discriminant $D(\vec{x})$ is a function of position within the variable space ${\cal V}$; at each point within that space $D(\vec{x})$ is a number between zero and unity, taking on values close to zero in regions that are predominantly populated by the types of collisions predicted by the hypothesis $B$, and taking on values that are close to unity in regions of ${\cal V}$ that are predominantly populated by the types of collisions predicted by the hypothesis $S$.

The set of all points $\vec{x}$ within ${\cal V}$ for which $D(\vec{x})>D_{\text{cut}}$ thus defines a region in the variable space ${\cal V}$ in which the contribution of collisions from the signal hypothesis $S$ is enhanced relative to the Standard Model background $B$.  Here $D_{\text{cut}}$ is some number between zero and one, chosen to optimize a figure of merit.  A common choice for this figure of merit is the number of events predicted by $S$ in the region divided by the square root of the number of events predicted by $B$ in the region, although more sophisticated choices for this figure of merit have been considered.

The number of events observed in the data within the region of ${\cal V}$ thus chosen is compared to the number of events predicted within that region by the ``background'' hypothesis $B$ and the ``signal + background'' hypothesis $S+B$, and from this comparison constraints are set on parameters internal to the hypothesis $S$, in some cases ruling out $S$ entirely.

One problem shared among all of these methods is that they are not sufficiently prescriptive.  The simple selection of a rectangular box in a multivariate space typically takes some fiddling, since a completely prescriptive procedure for choosing this box is lacking.  The parameters of a neural network typically have to be manipulated to achieve reasonable performance.  In the use of kernels, the choice of the width of each kernel and its covariance also lacks a sufficiently robust and general prescription, despite many proposals existing in the literature.  Obtaining a prescription that works reasonably and without pathologies in all cases would be incredibly helpful.

\subsection{Example: \Quaero}

An attempt at making the procedure sketched above completely prescriptive has been implemented in an algorithm called \Quaero~\cite{QuaeroPRL:Abazov:2001ny}.  Particle physics experimenters are notoriously guarded with their data, rarely sharing data in ``raw'' form with physicists outside their large collaborations.  The rationale for this is that the data in raw form are complicated by detailed effects and foibles of the detectors.  In order to perform a meaningful analysis of the data, the detector must be understood in much greater detail than anyone outside of the collaboration can reasonably hope to achieve.

This presents an obstacle to progress in particle physics, since the normal steps of testing any particular hypothesis within the collaboration (and then obtaining official collaboration approval of the result) typically takes greater than a year.  The solution to this conundrum is to package the knowledge of the physicists within the collaboration into a tool that can perform an analysis automatically --- a tool that ``knows'' how to correct for the foibles of the experimental apparatus.

\begin{figure}[tbh] \resizebox{3.5in}{!}{\hskip -0.5cm \includegraphics{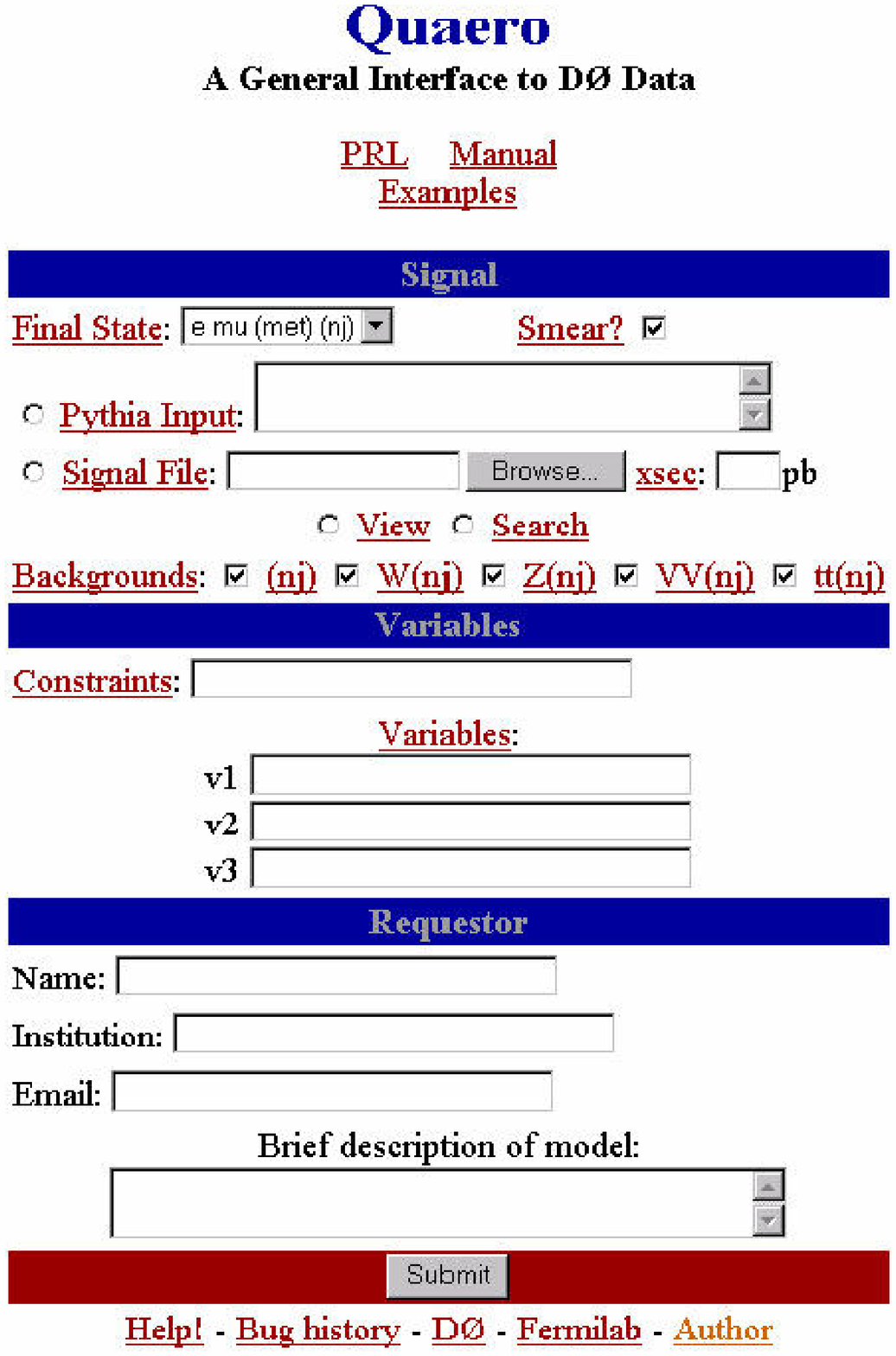} } \vskip -0.2in \caption{The \Quaero\ interface to D\O\ data.  A physicist outside the collaboration is able to provide his hypothesis in the form of the types of events he would expect should be observed were his hypothesis correct.  \Quaero\ performs the complete analysis, correctly accounting for expert knowledge of detector imperfections built into the algorithm, and returns to the requesting physicist constraints imposed by the data on the proposed hypothesis.  {\tiny{(Credit: the D\O\ Collaboration)}}} \label{fig:QuaeroWebPage} \end{figure}

Figure~\ref{fig:QuaeroWebPage} shows the \Quaero\ web page, through which the D\O\ collaboration has made a subset of its data publicly available.  A physicist can provide a file describing the type of events his model predicts should be seen in the detector, together with at most three variables that can be used to distinguish his signal ($S$) from the Standard Model ($B$).  \Quaero\ then constructs kernel density estimates $p(\vec{x}|S)$ and $p(\vec{x}|B)$ of the signal and background in the variables that the user has provided, uses these to construct the discriminant $D(\vec{x})$ according to Eq.~\ref{eqn:D}, chooses a number $D_{\text{cut}}$ to maximize a reasonable figure of merit, and from this determines the region of the variable space ${\cal V}$ for which $D(\vec{x})>D_{\text{cut}}$.  The number of events observed within that region in the data are then counted and compared to the number of events predicted by the physicist's signal and to the number of events predicted by the Standard Model alone.  From this comparison, parameters internal to the physicist's model are constrained.  Examples of the densities formed in this procedure are shown in Fig.~\ref{fig:Densities}.

\newcommand{\OnOffepsfbox}[1]
{\epsfxsize=3.3in\raisebox{-0.3in}[0.9in][0.3in]{\epsfbox{#1}}}
\begin{figure}[ht]
\centering
\begin{tabular}{ccc}
Background density & \ \ \ \ Signal density & Selected region \\ 
(a) & \ \ (b) & (c) \\ 
\multicolumn{3}{c}{\OnOffepsfbox{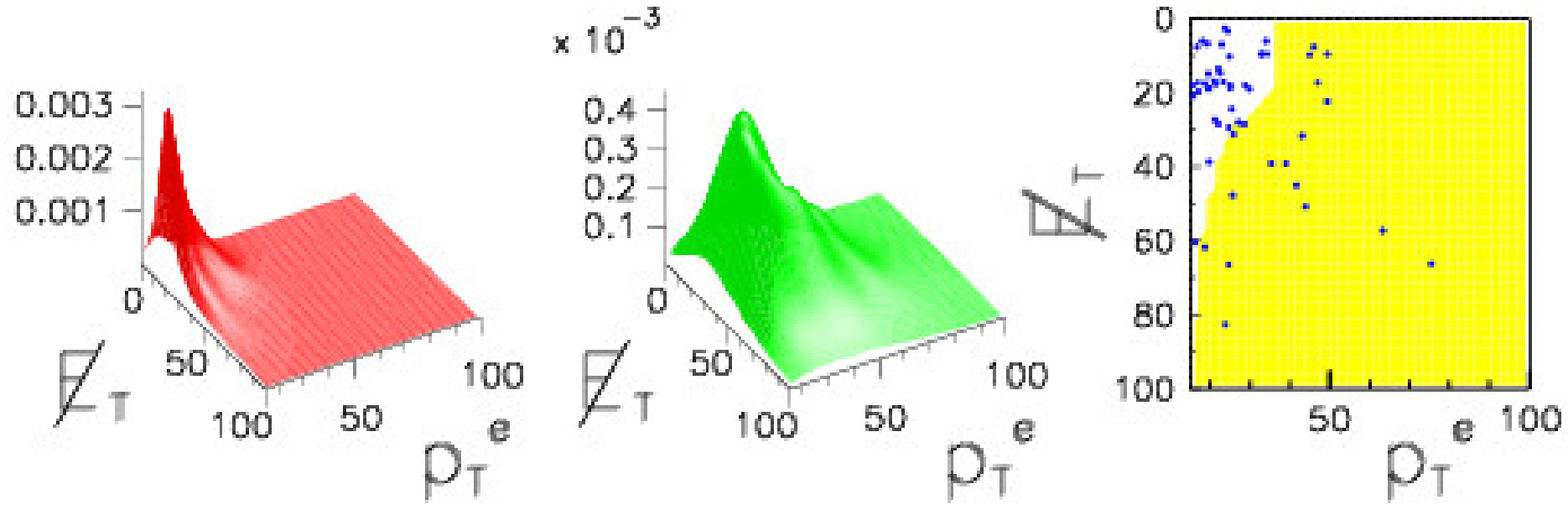}} \\
\multicolumn{3}{c}{\OnOffepsfbox{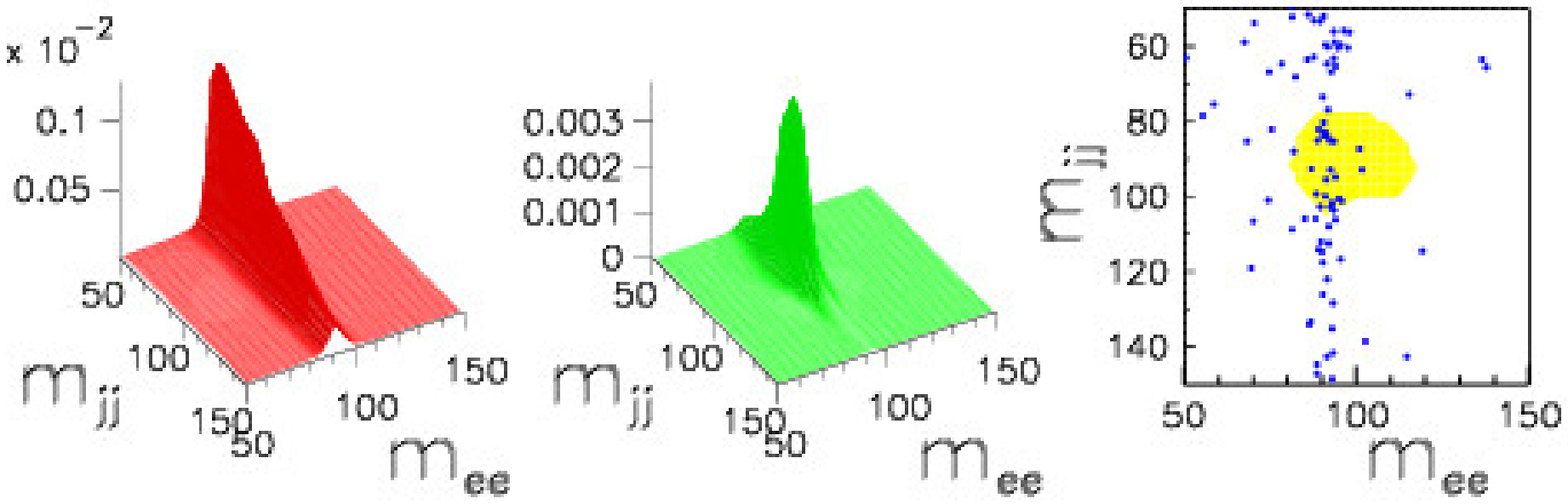}} \\
\multicolumn{3}{c}{\OnOffepsfbox{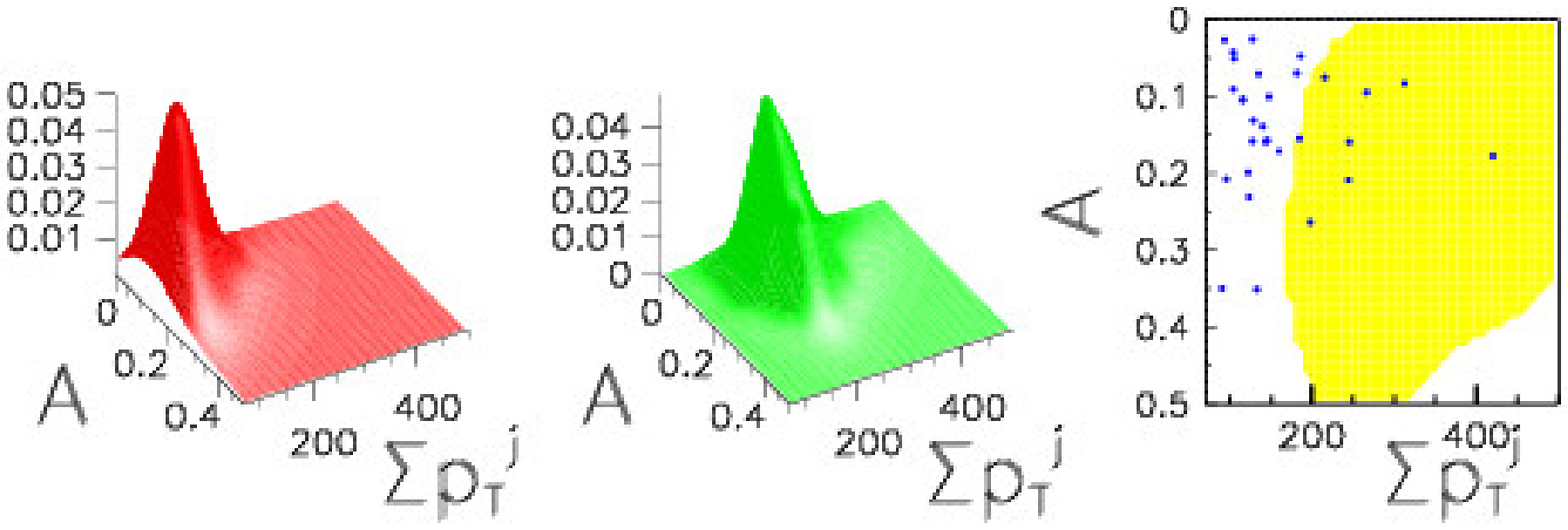}} \\
\multicolumn{3}{c}{\OnOffepsfbox{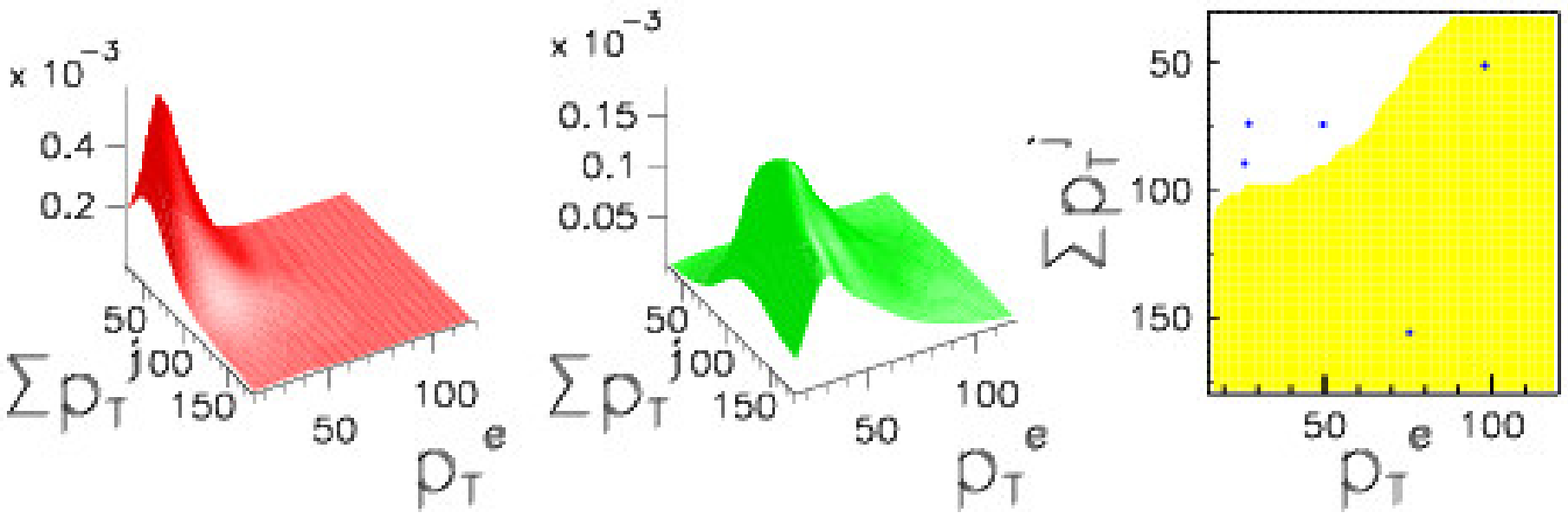}} \\
\end{tabular}
\caption{The background density (a), signal density (b), and selected region (shaded) (c) determined by \Quaero\ in several examples.  The dots in the plots in the rightmost column represent events observed in the data.  {\tiny{(Credit: the D\O\ Collaboration)}}}
\label{fig:Densities}
\end{figure}

\Quaero\ uses kernels whose shape is determined from the computation of a local covariance matrix, which appears to provide reasonable performance in most of the analyses that have been attempted.  It should be noted, however, that the reliance upon the ``correctness'' of the density estimate is in some sense minimal --- the density estimate is used only within a prescription to determine a particular region within the variable space ${\cal V}$ that enhances the contribution of signal relative to background.  Using the kernel density estimate directly to compute a likelihood requires confidence that the density estimate prescription used will yield non-pathologic results in all cases, and a prescription generating sufficient confidence is currently lacking.

By automating the testing of individual hypotheses, \Quaero\ dramatically decreases the time required to test models against large sets of data.  Each such test, formerly the subject of a Ph.D. theses, can now be done in an hour.

\section{\Sleuth\label{sec:Sleuth}}

Because \Quaero\ is designed to facilitate the testing of specific hypotheses against a large set of data, it requires that specific hypotheses be defined.  A separate algorithm is required to allow searches in the data for signatures of a more general type.  \Sleuth\ looks for anomalous deviations in the data that could indicate the presence of interesting physics, allowing the simultaneous testing of a large number of vaguely-specified hypotheses~\cite{KnutesonThesis,SleuthPRL:Abbott:2001ke,SleuthPRD1:Abbott:2000fb,SleuthPRD2:Abbott:2000gx}.

\subsection{The problem}

The Standard Model of particle physics has so far passed the experimental tests to which it has been subjected.  Nonetheless, there is a hole in the Standard Model that indicates there are likely to be new fundamental discoveries at energy scales that our accelerators are just beginning to probe.  Exactly what form that new physics might take is currently unknown, despite the work of hundreds of people over the last two decades.

Possible findings include 
\begin{itemize}
\item particles with bare magnetic charge (``magnetic monopoles''); 
\item a new symmetry of Nature (``supersymmetry'');
\item a new strong force (``technicolor'');
\item the presence of a new weak force (``heavy gauge bosons''); 
\item the existence of large extra spatial dimensions, curled up on scales smaller than 1 mm (``extra dimensions'');
\item electrons in excited states (``excited fermions'');
\item laws of nature that are not isotropic (``non-commutative theories'');
\item a heavier analog of the electron (``fourth generation of fermions''); and
\item evidence pointing to a unification of the strong, weak, and electromagnetic forces (``grand unified theories'').
\end{itemize}
Each of these possibilities represents a class of theories with a number of adjustable parameters.  When taken together, the model space is sufficiently large that systematically checking all possibilities is not a viable option.  We know only vaguely what it is we should be searching for; equivalently, we are searching for more things than can possibly be tested at one time.  

Humans are notoriously good at finding patterns, particularly when dealing with small numbers of events --- the history of particle physics is strewn with cases of patterns being mistakenly discerned.  An algorithm is required that will enable a general search for all possibilities simultaneously, rigorously taking into account the ``trials factor'' (a measure of the number of different possibilities considered) when reporting a final number.

\subsection{Solution}

The solution of this problem in the context of particle physics at accelerators that collide protons or their antiparticles is an algorithm called \Sleuth.  Consideration of the many possibilities just mentioned naturally leads one to ask whether there is any common feature among them.  If such a common feature exists, perhaps it can be searched for in a general way.

It turns out that such a commonality does in fact exist, justifying the following three assumptions:
\begin{itemize}
\item the data can be partitioned in such a way that a signal will predominantly appear within a single category;
\item interactions signaling the presence of new physics will generally produce objects with larger energy than expected from background processes; and
\item a signal is apt to appear as an excess of events --- i.e., more events observed in the data than expected from background.
\end{itemize}

\Sleuth\ begins by taking all of the data collected in the experiment and partitioning it into categories.  This partitioning is orthogonal; each event ends up in one and only one of these categories.  The partitioning is chosen such that if new physics appears in the data, it is likely to end up predominantly in a single category.  Each category contains a set of ``similar'' events, in the sense that the events in each category contain the same types of debris from the collision. 

Within each category, $d$ variables (where $d$ ranges between one and four, depending upon the category) are chosen.  In the case of \Sleuth, these variables correspond roughly to the energies of the objects produced in the collision.  The variables to be used for each category are set by a rule agreed upon before the data are analyzed.

\begin{figure}[tbh] \resizebox{3.5in}{!}{\hskip -0.5cm \includegraphics{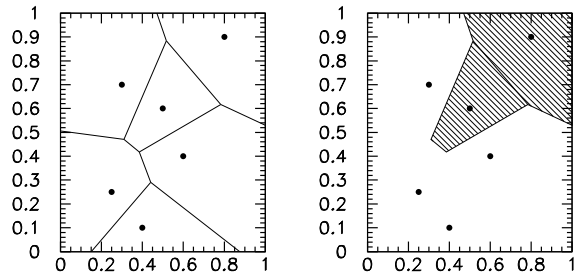} } \vskip -0.2in \caption{The seven black circles in each panel are data points within the unit square.  The Voronoi diagram (a) is formed by drawing perpendicular bisectors of imaginary line segments connecting neighboring pairs of points; the resulting lines partition the unit square into regions with the property that each space point inside the region is closer to the single data point inside that region than to any other data point in the square.  Regions considered by \Sleuth\ are unions of these individual regions, such as the shaded region in (b).  Criteria are imposed upon the regions that \Sleuth\ is allowed to consider, including the criterion that the region include the ``upper right-hand corner'' (1,1) of the unit square, as shown in (b).  {\tiny{(Credit: the D\O\ Collaboration)}}} \label{fig:VoronoiDiagram} \end{figure}

An arbitrary number of regions can be drawn in the variable space just defined.  In order to discretize the regions that can be considered, the following procedure is adopted.  A multivariate change of variables is made, such that in the new variables the prediction of the default hypothesis (the Standard Model) is a uniform distribution in $[0,1]^d$.  We refer to $[0,1]^d$ for arbitrary $d$ as the ``unit box,'' meaning unit interval when $d=1$, unit square when $d=2$, unit cube when $d=3$, and so forth.  Regions are then defined about sets of data points using the concept of Voronoi diagrams, an example of which is shown in Fig.~\ref{fig:VoronoiDiagram}.  The region around any single data point is the space within the unit box closer to that data point than to any other data point in the unit box.  The number of regions --- all possible unions of these individual regions --- that can be considered has been reduced by this process from uncountably infinite to a mere $2^{N_{\text{data}}}$, where $N_{\text{data}}$ is the number of events observed in the data in this category.   Because not all such regions are physically interesting, the regions considered are restricted to those containing the ``upper right-hand corner'' ($1^d$) of the unit box and satisfying other criteria favoring regions in which all variables take on values close to 1.  New physics is expected to appear in collisions with energetic particles in the final state; our mapping into the unit box is done such that these events will take on values close to the upper right hand corner.

The ``interestingness'' of each region is quantified as the Poisson probability that the background within that region will fluctuate up to or beyond the observed number of events.  A search heuristic is used to find the most interesting region $\scriptR$ in each category in the data.  The fraction $\scriptP$ of {\em hypothetical similar experiments}, in which a set of ``pseudo'' data points are drawn from the background distribution, that give rise to a result more interesting than the most interesting region observed in the data is determined.  $\scriptP$ is a number between zero and unity; $\scriptP$ will be a (uniformly-distributed) random number between zero and unity if the data are composed of background only; $\scriptP$ should be close to zero if there really is something interesting in the data.  One value of $\scriptP$ is determined for each category.  The minimal $\scriptP$ in the (roughly 100) categories considered determines $\twiddleScriptP$, the fraction of hypothetical similar experiments in which a more interesting region would be found in any of the categories considered.  This $\twiddleScriptP$ is the final measure of ``interestingness.''  If the data are distributed according to background only, $\twiddleScriptP$ will be a number randomly distributed in the unit interval.  If the data contain a hint of a signal, $\twiddleScriptP$ will (hopefully) be small.  The output of \Sleuth\ is thus twofold:  the most interesting region $\scriptR$ in any of the categories considered, and a quantitative measure $\twiddleScriptP$ of how interesting that region actually is.

\subsection{Results}

\begin{figure}[tbh] \resizebox{3.5in}{!}{\hskip -0.5cm \includegraphics{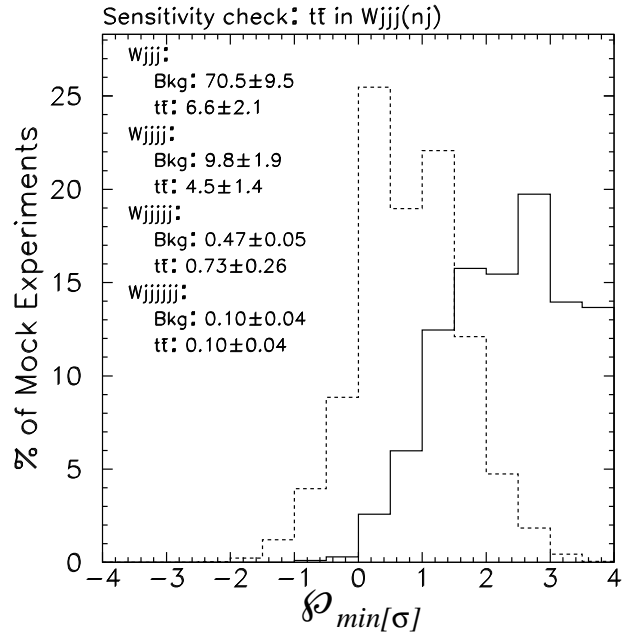} } \vskip -0.2in \caption{Examples of \Sleuth's performance on a number of mock experiments for a particular signal (denoted $t\bar{t}$).  This particular signal manifests itself in four categories (here labeled $Wjjj$, $Wjjjj$, $Wjjjjj$, and $Wjjjjjj$).  The number of ``background'' (Bkg) events predicted by the Standard Model and by the $t\bar{t}$ signal is shown in the legend.  The horizontal axis is the minimum $\scriptP$ found in the four categories, in units of standard deviation; the vertical axis is the fraction of mock experiments in which this value of $\scriptP$ is obtained.  The dashed line is the minimum $\scriptP$ found when the data in the mock experiments are pulled from the background distribution; the solid line is the minimum $\scriptP$ found when the data in the mock experiments are pulled from the background with an admixture of the $t\bar{t}$ signal.  {\tiny{(Credit: the D\O\ Collaboration)}}} \label{fig:WjjjxTop} \end{figure}

By construction, \Sleuth\ will not find anything if there is nothing there to be found.  But would \Sleuth\ find something if there were something there to be found?  This is impossible to address adequately in the general case; the answer depends to what degree the underlying physical assumptions made in \Sleuth\ are realized in Nature.  The question can, however, be definitively answered in any specific scenario, and several studies have been performed in order to test \Sleuth's sensitivity to a variety of new physics.  The results of one such example are shown in Fig.~\ref{fig:WjjjxTop}.  Other studies in various categories with different signals indicate that \Sleuth\ performs as anticipated, the success of the algorithm in each case depending upon the strength of the signal and the extent to which the signal satisfies the three assumptions on which the algorithm is based.  Such studies indicate that \Sleuth\ has a reasonable shot of discovering new physics were it present in our data.

\begin{figure}[tbh] \resizebox{3.5in}{!}{\hskip -0.5cm \includegraphics{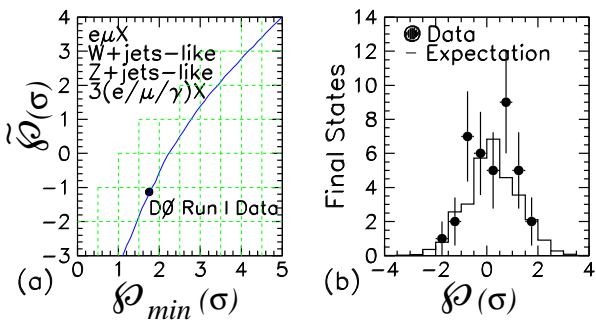} } \vskip -0.2in \caption{For each of the categories considered at D\O, one value of $\scriptP$ was determined.  These values, in units of standard deviation, are histogrammed as filled circles in (b); the solid line shows the distribution expected if the data consist of background only.  The minimum $\scriptP$ found in all categories considered must be mapped to the final output $\twiddleScriptP$ by taking into account the many categories that have been considered.  The horizontal axis in (a) represents the minimum $\scriptP$ found, in units of standard deviation; the vertical axis represents the final output $\twiddleScriptP$, also in units of standard deviation; the blue curve shows the relationship between the minimum $\scriptP$ found and the corresponding final output $\twiddleScriptP$.  At D\O, the minimum $\scriptP$ was found to be 0.04; from (a), this is seen to correspond to $\twiddleScriptP = 0.89$.  {\tiny{(Credit: the D\O\ Collaboration)}}} \label{fig:ResultsPlotPRL} \end{figure}

\Sleuth\ has been applied to a range of particle collider data collected between 1992--1996 by the D\O\ experiment at the Fermilab Tevatron.  Disappointingly, a null result has been obtained.  The $\scriptP$'s obtained in thirty-two different categories are histogrammed in Fig.~\ref{fig:ResultsPlotPRL}(b), and the conversion to $\twiddleScriptP$ given the number of categories considered is shown in Fig.~\ref{fig:ResultsPlotPRL}(a).  

\begin{figure}[tbh] \resizebox{3.5in}{!}{\hskip -0.5cm \includegraphics{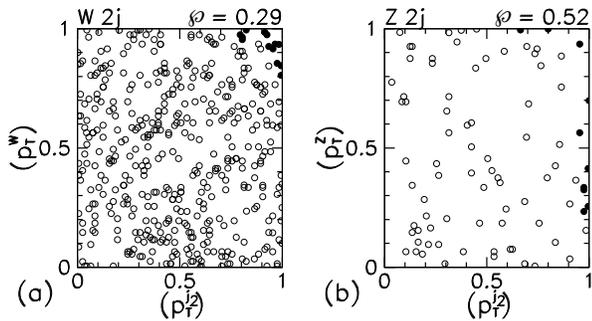} } \vskip -0.2in \caption{Examples of \Sleuth's evaluation of data in two categories (denoted $W\,2j$ and $Z\,2j$).  The open circles are data points outside the region selected by \Sleuth; the filled circles are data points inside the region selected by \Sleuth.  The value of $\scriptP$ determined in each case is shown upper right.  {\tiny{(Credit: the D\O\ Collaboration)}}} \label{fig:ResultsDataPlotPRL} \end{figure}

\subsection{Remarks}

The general problem of looking for a signal that is only vaguely specified is of course not peculiar to particle physics, and the general thread of the solution should therefore be applicable to any problem of similar type.  The individual steps of the algorithm --- partitioning the data into categories, defining a small set of variables for each category, rigorously defining regions within each variable space, specifying criteria for the regions to be considered, quantifying the interestingness of any particular region, searching the data to find the most interesting region within each category, conducting hypothetical similar experiments to determine $\scriptP$ for each category, and conducting a second set of hypothetical similar experiments to determine $\twiddleScriptP$ for the data as a whole --- are generally modifiable to many problems.  Where the size of the data set is an issue, a specific solution can often be found that scales with the number of events $n$ as ${\cal O}(n\log n)$.

\section{Summary and Discussion}

The massive data sets from this decade's particle physics experiments present a variety of challenges.  The selection of which collisions to record on permanent storage (triggering), discussed in Sec.~\ref{sec:Triggering}, presents a need for smart algorithms whose implementation in a parallel hardware architecture can take the high rate of incident data.  Understanding and optimizing the necessary trade-offs imposed by the constraints of each system involve design patterns that are still being learned.  Imposing deadlines and ignorance of theoretical best practice result in systems that can be noticeably suboptimal.

The reconstruction of the trajectories of the primary components of the debris emerging from the collision, sketched in Sec.~\ref{sec:Reconstruction}, involves a number of pattern recognition problems that can be handled at the more leisurely pace of one second per collision.  Track finding, a connect-the-dots problem of identifying curved particle trajectories in the presence of noise and complicating interactions of the particle with the traversed material, is prototypical problem.  Tracking flying objects with sweeping radar or swimming objects emitting infrequent signals are two of many examples where similar issues are encountered.  Jet clustering, the clustering of the debris observed in the detector, has similarities to many of the clustering problems currently treated in the statistics literature.  These are two areas in which the statistical community's impact on current practice in our field has been significantly less than it could be. 

Kernel Density Estimation has been recognized by the field as a useful technique for estimating parent densities of finite samples, the limited statistics of which are often imposed by the time required to simulate what would be observed in our detectors if a particular model were realized in nature.  \Quaero, the intelligent web interface to D\O\ proton anti-proton collision data described in Section~\ref{sec:KernelDensityEstimation}, serves as an existence proof that the testing of particular hypotheses against high energy collider data can be formalized and automated.  Outstanding issues here include whether a completely prescriptive kernel density algorithm can be achieved that avoids pathologies when presented with outliers, delta functions in an otherwise continuous distribution, and discontinuous boundaries.  To be a bit more provocative: For every existing kernel estimation prescription of which we are aware, a sample exists for which the prescription produces a clearly undesirable estimate of the parent density.  Presumably a prescription that is never obviously unreasonable can be achieved, and this would be of great value.  Such a prescription whose time cost grows only linearly with the size of the sample (even in the multidimensional case) would be of even greater value.

The problem of how to look for a vaguely specified feature in a large set of data, discussed in Section~\ref{sec:Sleuth}, is omnipresent in scientific research.  Having the ability to rigorously quantify the ``interestingness'' of a particular subset of collisions in an unbiased manner is found to be crucial.  This basic lesson can be widely generalized.  The solution (\Sleuth) is obtained by constructing a rigorous quantification of ``interestingness'' before the data is analyzed, so that pseudo experiments can be simulated to determine what fraction of them would yield data as interesting as what is actually observed.  Obvious applications range from identifying statistically significant galaxy clusters to identifying statistically significant disease clusters.

Opportunities for the computational and graphical statistics community to contribute to the solutions of problems faced in particle physics abound; this article has barely scratched the surface.  We look forward to a productive exchange of ideas over the coming decade.

\acknowledgments
P.~P.~thanks David Scott for the invitation to a lively conference at which this talk was given, and his colleagues at CMS and D\O\ for the collaborative work from which the discussion of triggers in Sec.~\ref{sec:Triggering} has been drawn.  B.~K.~thanks his colleagues in the D\O\ collaboration, discussions with whom led to the development of \Quaero\ and \Sleuth, and his collaborators on CDF.  P.~P.~is supported by a grant from the Department of Energy; B.~K.~is supported by a Fermi/McCormick Postdoctoral Fellowship at the University of Chicago.

\bibliography{jcgs}

\end{document}